\documentclass[11pt, a4paper]{article}

\usepackage[left=2.5cm,
            right=2.5cm,
            top=2.5cm,
            bottom=2.5cm]{geometry}
            \usepackage[T1]{fontenc}
\usepackage[utf8]{inputenc}
\usepackage{graphicx, xcolor}
\usepackage{lmodern}
\usepackage{amsmath, amssymb, mathtools}
\usepackage{longtable, booktabs, array}
\usepackage{orcidlink}
\usepackage{setspace}

\usepackage[]{natbib}
\makeatletter
\newcommand\code{\bgroup\@makeother\_\@makeother\~\@makeother\$\@codex}
\def\@codex#1{{\ttfamily\hyphenchar\font=-1 #1}\egroup}
\makeatother

\newcommand{\J}{\boldsymbol{J}}

\newcommand{\Normal}{\mathcal{N}}

\renewcommand{\hat}[1]{\widehat{#1}}

\newcommand{\mub}{\boldsymbol{\mu}}
\newcommand{\Sigmab}{\boldsymbol{\Sigma}}

\begin{document}

\title{\LARGE\bfseries A Model-Based Approach to\\ Shot Charts Estimation in Basketball}
\author{
\large Luca Scrucca~\orcidlink{0000-0003-3826-0484}\\
\normalsize Department of Economics, Università degli Studi di Perugia\\[1ex]
\large Dimitris Karlis~\orcidlink{0000-0003-3711-1575}\\
\normalsize Department of Statistics, Athens University of Economics
}
\date{\today}

\maketitle

\bigskip
\begin{abstract}
\noindent%
Shot charts in basketball analytics provide an indispensable tool for evaluating players' shooting performance by visually representing the distribution of field goal attempts across different court locations. However, conventional methods often overlook the bounded nature of the basketball court, leading to inaccurate representations, particularly along the boundaries and corners. 
In this paper, we propose a novel model-based approach to shot chart estimation and visualization that explicitly considers the physical boundaries of the basketball court. By employing Gaussian mixtures for bounded data, our methodology allows to obtain more accurate estimation of shot density distributions for both made and missed shots. Bayes' rule is then applied to derive estimates for the probability of successful shooting from any given locations, and to identify the regions with the highest expected scores. 
To illustrate the efficacy of our proposal, we apply it to data from the 2022-23 NBA regular season, showing its usefulness through detailed analyses of shot patterns for two prominent players.
\end{abstract}

\tableofcontents	

\clearpage

\section{Introduction}

Basketball is among the most popular sports game worldwide. It not only enjoys widespread popularity as a sport but has also generated substantial economic benefits through its associated industries.
The National Basketball Association (NBA) is widely recognized as the world’s leading league, attracting international interest with tremendous amounts spent in related marketing.
In Europe, the Euroleague represents the pinnacle of professional men's club basketball competition and is regarded as the top-tier men's league on the continent. 
The increasing interest on basketball has led quite early to the development of
advanced statistical methodologies for measuring performance \citep{kubatko2007starting}, while several other proposal have been made after this.
For a broad picture of academic and non-academic research on basketball analytics we recommend the book of \cite{zuccolotto2020basketball} and the broad review paper by \cite{terner2021modeling}.

One of the basic characteristics of basketball is that it is a fast-paced
contact game in which the players are constantly moving in heated confrontations, thus leading to quick transitions from defense to offense or vice versa. 
In practice basketball is a game of space \citep[see e.g.][p1]{goldsberry2012courtvision}. The teams that makes better use of spatial aspects can have an advantage and hence several tactics related to better enhancement of spatio-temporal game aspects \citep{sandholtz2020measuring}.

Advancements in sports information systems and technology has allowed the collection of a number of detailed spatio-temporal data that capture various aspects of basketball \cite{papalexakis2018thoops,shortridge2014creating}. Such data can help considerably to understand the game and the effects of space on that while they also provide interesting information for all stakeholders of the game, including trainers, team managers, players, scouters of new players, spectators and journalists.
Visualizations of basketball games can provide important information about the game \citep{perin2018state}. 
An increasing number of visualization research has been conducted that includes as visual analysis of player trajectories, visualization of field goals of a player, and visualization of basic statistics of different players in different games \citep{chen2016gameflow}.

Shots are a key-ingredient of the sport. 
The final score of a team is defined by the number of successful shots and their quality, (2 or 3 points plus the 1 point for free throws). 
As such considerable interest has been made on understanding and predicting shot tactics and success.
For example, \cite{zuccolotto2018big} utilized several techniques to model scoring probability under high-pressure conditions in basketball based  play-by-play data from the Italian “Serie A2” Championship 2015/2016.
\cite{shortridge2014creating}  discussed and proposed different measures about shot efficiency that take into account the spatial effect and they also proposed visualizations related to shot efficiency. 
\cite{oughali2019analysis} tried to predict shot success based on several machine learning approaches. 
\cite{fichman2019optimal} discussed the optimal shot selection strategy for a basketball team.
\cite{jiao2021bayesian} proposed a marked spatial point process for modeling 
basketball shots based on the observation that the success rate of a basketball shots may be higher at locations where a player makes more shots. Related to the spatial aspect are also the so-called corner 3's, which are those shots that while producing 3 points are taken closer to the basket, thus allowing for larger probability of success and distinguished tactic for that shots \citep{pelechrinis2021anatomy}.

Visualizing shots can be a powerful tool for better understanding the different tactics. 
Quite early it has been noted that spatial visualizations like shot charts can be very valuable to reveal the tactical performance of the teams and hence be a valuable tool in the hands of trainers \citep{reich2006spatial}. 
For example, shot charts, that is, maps capturing locations of (made or missed) shots, and spatio-temporal trajectories for the players on the court can capture information about the offensive and defensive tendencies, as well as, schemes used by a team. Characterization of these processes is important for player and team comparisons, scouting, game preparation etc.
Since then there has been extensive literature related to shots in basketball including effective visualizations that can produce insights. 
Since shots are the most important aspect as it leads to gaining points, it is quite common to produce statistics related to shot success but also to shot patterns, including spatio-temporal aspects of shots. 
The radical choice of most teams towards different shooting styles that include more 3-points is perhaps partially due to the improved visualizations available. 

\emph{Shot charts} in basketball analytics are a fundamental tool for visually examining the distribution of players' field goal attempts and their efficiency in different court locations.
Typically such charts visualize the locations of all shots made, either by cutting the courts in cells (or hexagons or other areas) of the same size and representing their frequency by some color \citep{chu2010information}. \cite{ehrlich2024estimating} proposed alternative ways to improve the information provided using some model based estimate of the shot efficiency. See also the work on \cite{fu2024hoopinsight} about the importance of visualizing the shooting performance. 

However, despite their utility, current shot charts representations face certain limitations. Predominantly, when constructed from observed data using hexagons or derived from standard density estimation procedures, they often fail to take into account the bounded nature of the basketball court. This limitation can result in misleading representations, especially at the boundaries and corners of the court.
Consequently, the analysis may not fully account for the contextual constraints imposed by the court's physical boundaries, potentially skewing the assessment of shooting patterns and efficiency, particularly in areas where players are more inclined to attempt shots due to strategic advantages or positional play. These discrepancies underscore the importance of refining shot charts methodologies to accurately depict the nuanced spatial dynamics inherent in basketball shot data.

\begin{figure}[htb]
\centering
\includegraphics[width=\textwidth]{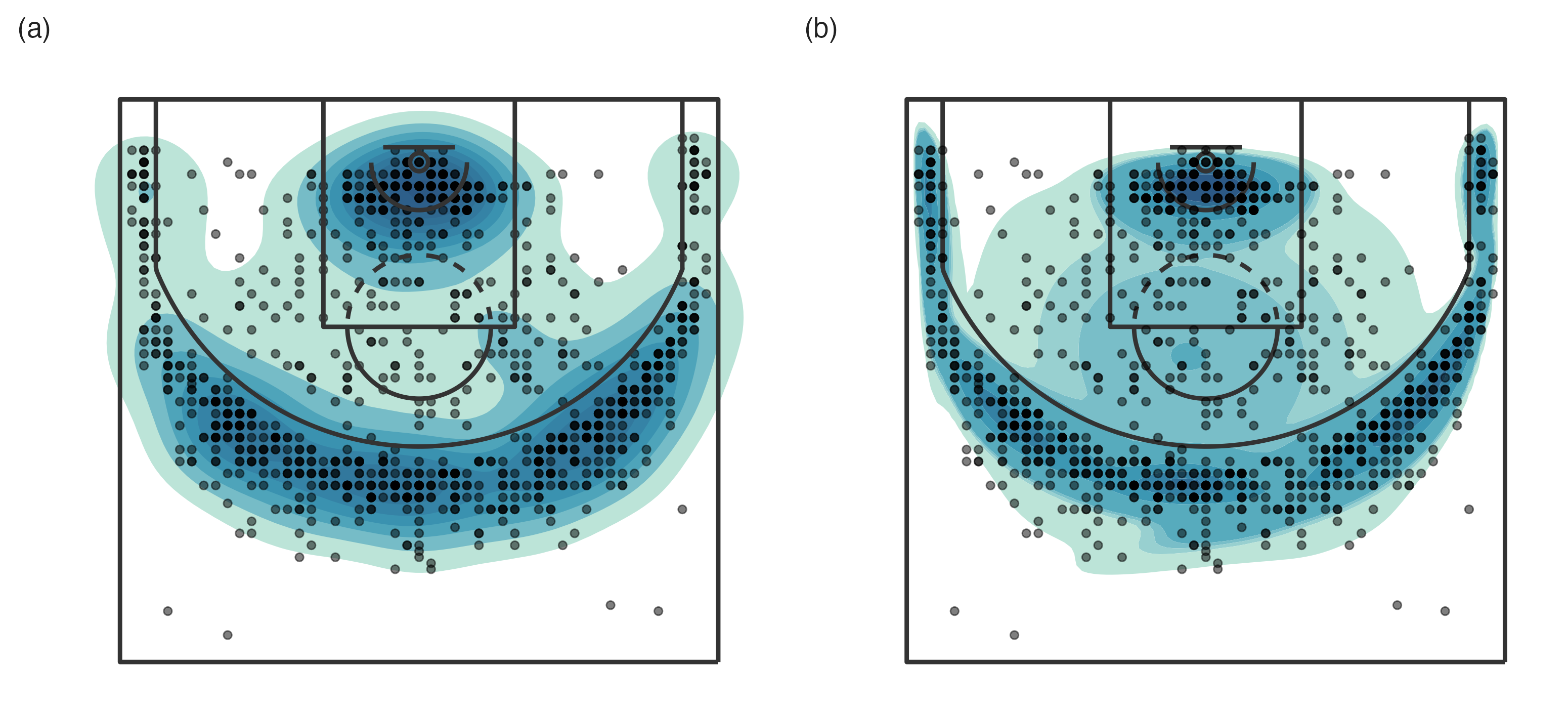}
\caption{Distribution of Stephen Curry's shot attempts during the 2022-23 NBA regular season. Panel (a) shows the density estimate obtained using two-dimensional kernel density estimation, while panel (b) the estimate obtained by fitting Gaussian mixtures for bounded data, which allows the physical boundaries of the basketball court to be taken into account.}
\label{fig:shotchart-density}
\end{figure}

Figure~\ref{fig:shotchart-density}a illustrates the approach commonly used to visualize the spatial distribution of a player's shot attempts. Typically, a two-dimensional kernel density estimate is used \citep{Scott:2009}. However, if the boundaries of a basketball court are not taken into account, some artifacts are noticeable, particularly in the corner 3-point areas, behind the backboard, and in front of the center 3-point line.
In contrast, by adopting the methodology proposed in this paper we obtain a density estimate that remains confined within the physical boundaries of the basketball court and, by providing more accurate spatial estimates, effectively remove the above mentioned artifacts.

To summarize, in our proposal we embrace a model-based approach to shot charts estimation and visualization that:
1) employs Gaussian mixtures to estimate the density distribution of made and missed shots;
2) takes into account the physical boundaries of the basketball court;
3) applies Bayes' rule to derive estimates for the probability of successful shooting from any location; and
4) identifies regions with the highest expected scores.

The paper is organized as follows: Section~\ref{sec:methods} describes the model and the estimation procedure; Section~\ref{sec:applications} illustrates the proposed methodology using the data from the 2022-23 NBA regular season for two players, namely Stephen Curry, perhaps the GOAT (\emph{Greatest Of All Time}) 3-point shooter, and Joel Embiid, the MVP (\emph{Most Valuable Player}) for that season; the final section contains some concluding remarks and potential future extensions to this paper.

\section{Methods}
\label{sec:methods}

Shot charts in basketball analytics provide a visual representation of a player or team's shooting performance by analyzing data on shots attempted from various spots on the court. 
However, basketball courts come in many different sizes. In the NBA, the court is 94 by 50 feet (28.7 by 15.2 m), while under the International Basketball Federation (FIBA) rules, the court is slightly smaller, measuring 28 by 15 meters (91.9 by 49.2 ft). 
The 3-point line is also different, being located at 23 feet 9 inches (7.24 m) from the center of the basket in the NBA (22 ft or 6.70 m at the corner), and 6.75 m (22 ft 1.75 in) for FIBA (6.60 m or 21 ft 8 in at the corner).
As discussed in reference to the results shown in Figure~\ref{fig:shotchart-density}, these physical constraints on the basketball court must be given due consideration in density estimation from shots spatial information.

\begin{figure}[htb]
\centering\bigskip
\includegraphics[width=0.48\textwidth]{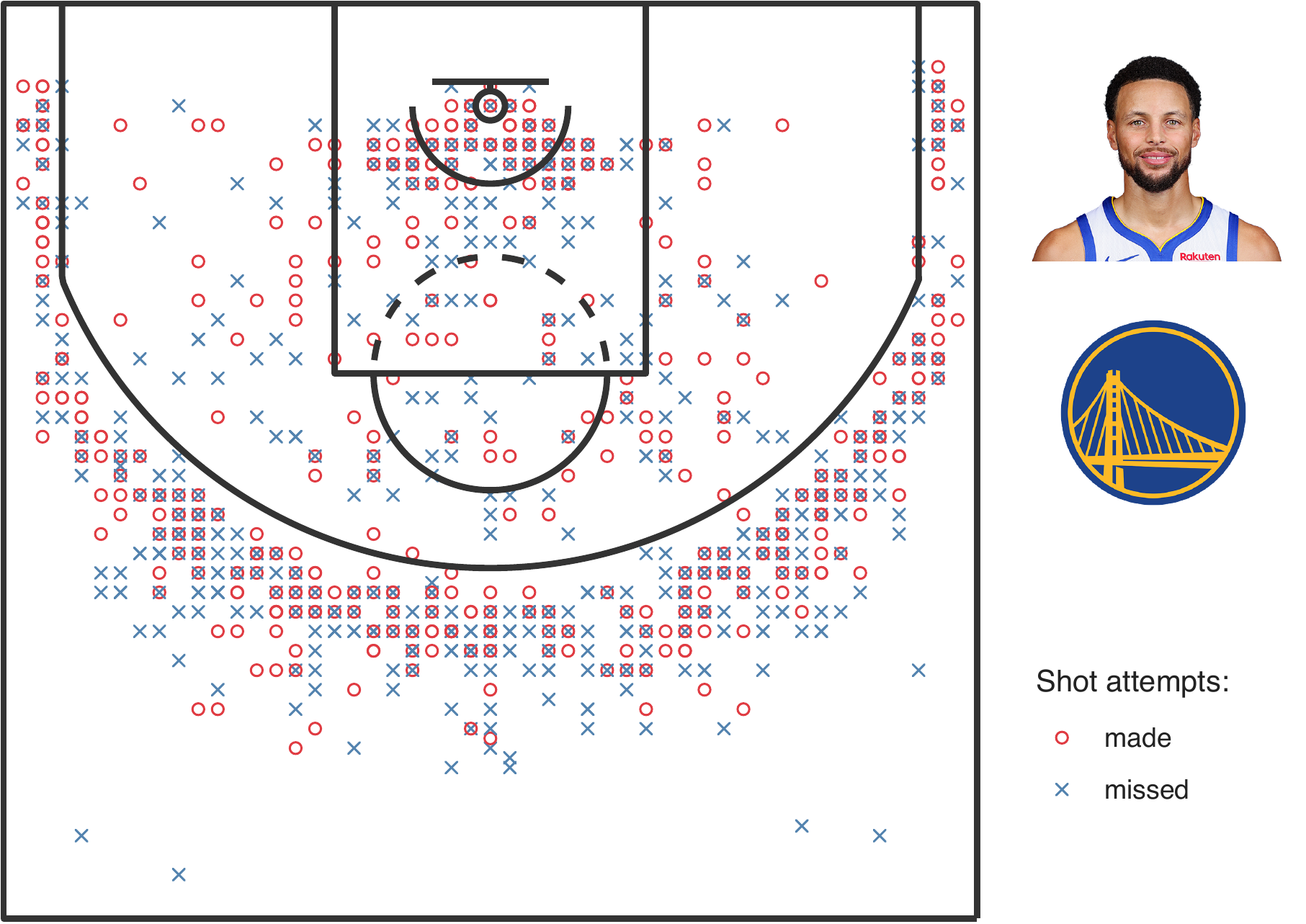}\hfill
\includegraphics[width=0.48\textwidth]{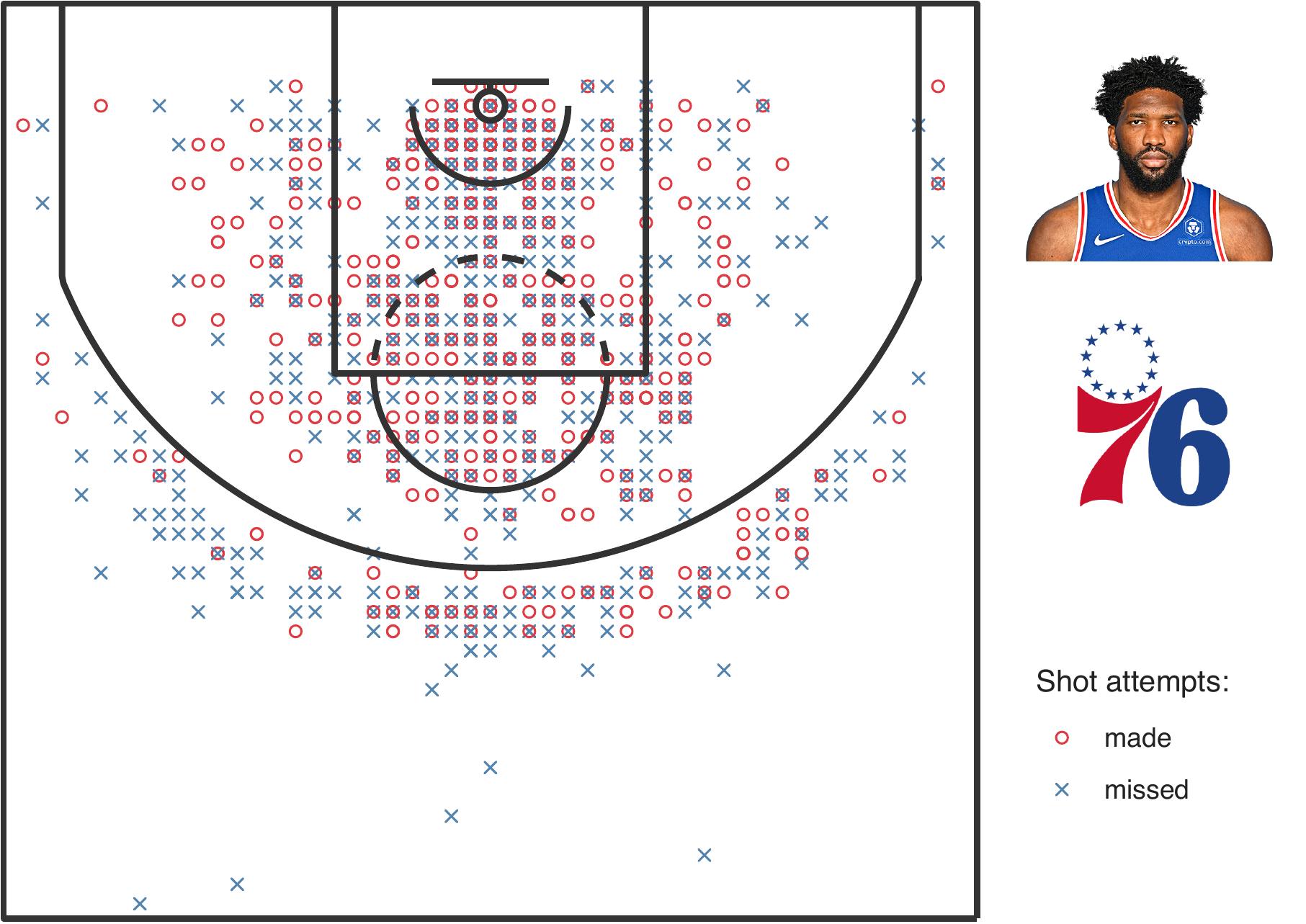}
\caption{Shots attempted by Stephen Curry and Joel Embiid during the 2022-23 NBA regular season.}
\label{fig:curry-embiid-shots}
\end{figure}

Figure~\ref{fig:curry-embiid-shots} shows the shots attempted by Stephen Curry (left panel) and Joel Embiid (right panel) during the 2022-23 NBA regular season with each data point marked by shot outcome.
The significant presence of shots from beyond the arc of the 3-point line is evident for Curry, while a greater number of attempts in the mid-range can be traced for Embiid. However, partly because of the presence of overlapping points, it is difficult to identify the spots from which the two players preferentially and most effectively shoot at the basket. 
Thereby, density estimation becomes crucial for gaining insights into shooting patterns and optimizing players performance, or to set up an efficient defense that limits shooting opportunities at preferred positions.

Gaussian mixtures \citep{McLachlan:Peel:2000,Fraley:Raftery:2002} offer a semiparametric approach to density estimation.
In this approach, the density of the data is expressed as a convex linear combination of one or more probability density functions. Gaussian mixtures are a popular choice obtained by using Gaussian densities as components of the mixture.

Gaussian Mixtures Models (GMMs) carry several advantages due to their intrinsic probabilistic generative nature. In particular, maximum likelihood estimation of parameters is available via the EM algorithm (see Section~\ref{sec:estmodsel}), with estimates that remain efficient even for multidimensional data. Moreover, GMMs require no hyperparameters tuning, with the problem of selecting the complexity of the mixture that can be recast as model selection problem (see Section~\ref{sec:estmodsel}).

Despite the fact that GMMs can approximate any continuous density with arbitrary accuracy, provided the mixture has an adequate number of components \citep[see][among others]{Ferguson:1983, Escobar:West:1995}, it is crucial to consider the inherent physical constraints of the basketball half-court when estimating densities in shot charts.
This can be achieved by adopting the transformation-based approach to Gaussian mixture density estimation for bounded data proposed by \citet{Scrucca:2019}. 
This approach is particularly suitable for this scenario because it explicitly considers the natural bounds of the basketball half-court.
Next section briefly reviews the methodology of our proposal.

\subsection{Model specification}
\label{sec:modspec}

The transformation-based approach for GMMs discussed in \citet{Scrucca:2019} extends density estimation using mixture modeling to the case of bounded variables. The basic idea is to carry out density estimation not on the original data but on appropriately transformed scale. Then, the density for the original data can be simply obtained by a change of variables. 

\begin{figure}[ht]
\centering
\includegraphics[width=0.6\textwidth]{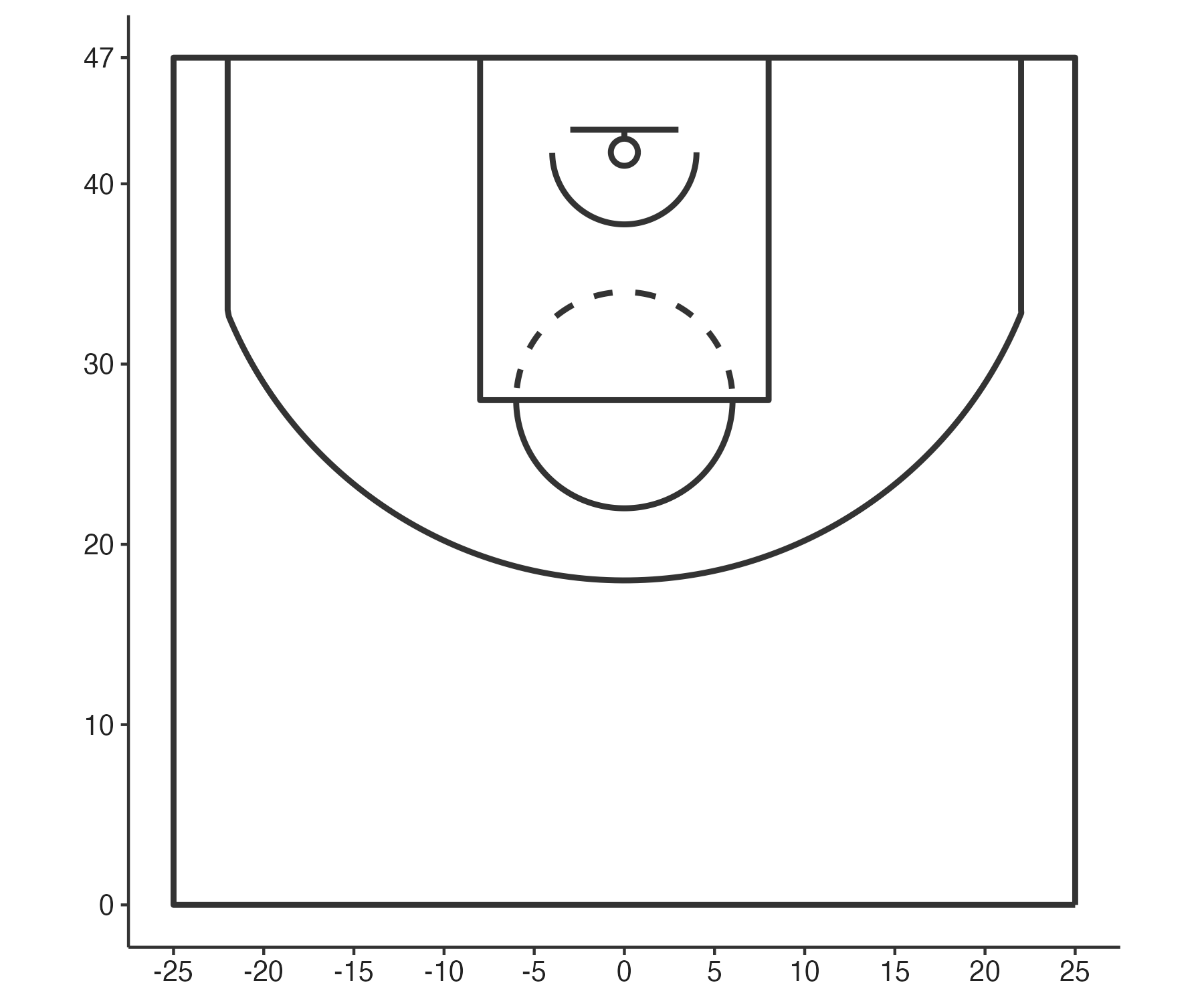}
\caption{NBA half-court dimensions and coordinates (in feet) used in the present paper.}
\label{fig:basketball-court}
\end{figure}

Let $(x_i, y_i)$ denote the coordinates of the position on the court where a player attempts a shot, for $i = 1, \dots, n$, where $n$ is the number of shots attempted, and $C_i = \{ 0, 1 \}$ the corresponding binary outcome, where $1$ indicates a made shot and $0$ a missed shot.
Consider the coordinate-wise range-logit transformation defined as
$$
t(x,y) = 
\begin{bmatrix}
t(x)\\
t(y)
\end{bmatrix} =
\begin{bmatrix}
\log\left( \frac{x-\ell_x}{u_x - x} \right) \\[1ex]
\log\left( \frac{y-\ell_y}{u_y - y} \right) \\
\end{bmatrix},
$$
where $(\ell_x, u_x)$ and $(\ell_y, u_y)$ are the lower and upper bounds along, respectively, the $x$-axis and the $y$-axis. 
Figure~\ref{fig:basketball-court} shows the coordinates of the half-court we consider in our study for a 94 by 50 feet NBA basketball court. Thus, half-court court boundaries are set at $(\ell_x = -25, u_x = 25)$ and $(\ell_y = 0, u_y = 47)$.

In the logit-range transformed scale the density of a shot from location $(x,y)$ can be expressed using the following Gaussian mixture 
\begin{equation}
h(t(x,y)) = \sum_{g = 1}^G \pi_g \;\Normal(t(x), t(y) \mid \mub_{g}, \Sigmab_{g}),
\label{eq:mixdens}
\end{equation}
where $G$ is the number of mixture components, $\pi_{g}$ the mixing probabilities
(with $\pi_{g} > 0$ and $\sum_{g=1}^{G}\pi_{g}=1$), $\mub_{g}$
and $\Sigmab_{g}$, respectively, the mean vector and covariance matrix for Gaussian component $g$.
Upon re-expressing it in the original coordinate scale, the density function can be formulated as follows:
\begin{equation}
f(x,y) = h(t(x,y)) \times |\J(t(x,y))|,
\label{eq:xydensfun}
\end{equation}
where $|\J(t(x,y))|$ is the Jacobian of the transformation. 
According to the coordinate-wise transformation approach adopted, the matrix of first derivatives is diagonal, so the Jacobian reduces to the product of first derivatives, i.e.
$$
|\J(t(x,y))| = t'(x) \times t'(y) = 
\left( \frac{1}{x-\ell_x} + \frac{1}{u_x-x} \right) \times
\left( \frac{1}{y-\ell_y} + \frac{1}{u_y-y} \right).
$$

The density in the transformed coordinates from \eqref{eq:mixdens} can be estimated separately for made ($C=1$) and missed shots ($C=0$), and then back-transformed in the original scale using \eqref{eq:xydensfun}. 
Subsequently, the probability of scoring a basket from a specific location can be calculated using Bayes' theorem.
Specifically, the density at location $(x,y)$ for shot outcome $C=k$, with $k=\{0,1\}$, is given by 
\begin{equation}
f(x, y \mid C = k) = 
\left( 
  \sum_{g=1}^{G_k} \pi_{g|k} \;\Normal(t(x), t(y) \mid \mub_{g|k}, \Sigmab_{g|k})
\right)
\times |\J(t(x,y))|,
\label{eq:xydens}
\end{equation}
where $G_k$ represents the number of mixture components for shot outcome $C=k$. 
The $\pi_{g|k}$ terms denote the mixing probabilities for outcome $C=k$ ($\pi_{g|k} > 0$ and $\sum_{g=1}^{G_k}\pi_{g|k}=1$), and $\mub_{g|k}$ along with $\Sigmab_{g|k}$ stand for the mean vectors and covariance matrices for component $g$ of outcome $C=k$.

Once the density is estimated for both made shots, $f(x, y \mid C = 1)$, and missed shots, $f(x, y \mid C = 0)$, the probability of a successful shot can be obtained using Bayes' rule as:
\begin{equation}
\Pr(C = 1 | x, y) = 
\frac{\tau_1 f(x, y \mid C = 1)}{ \tau_0 f(x, y \mid C = 0) + \tau_1 f(x, y \mid C = 1)},
\label{eq:xycondprob}
\end{equation}
where $\tau_1$ and $\tau_0$ are the outcome prior probabilities of, respectively, made and missed shots.

The estimated probabilities of making shots from various positions on the court in \eqref{eq:xycondprob} can be multiplied by the point value of those shots (2 or 3 points) to derive the \emph{expected points scored}:
$$
\text{EPS}(x,y) = 
\begin{cases}
2 \times \Pr(C = 1 | x, y) & \text{if $(x,y)$ is within the 3-point line} \\
3 \times \Pr(C = 1 | x, y) & \text{if $(x,y)$ is beyond the 3-point line}
\end{cases}
$$
This represents an important metric which provides valuable insights into offensive strategies and efficiency from different positions on the court.

\subsection{Estimation and model selection}
\label{sec:estmodsel}

Estimation of unknown parameters, $\pi_{g|k}, \mub_{g|k}, \Sigmab_{g|k}$, for $g=1,\ldots,G_k$ and $k=\{0,1\}$, in \eqref{eq:xydens} can be pursued via the EM algorithm. For details see \citet[Sec. 3.3]{Scrucca:2019}. Moreover, outcome prior probabilities, $\tau_1$ and $\tau_0$, in \eqref{eq:xycondprob} can be estimated from, respectively, the proportions of made and missed shots.

Without imposing any constraints on the covariance matrices of Gaussian components, empirical evidence suggests the inclusion of a Bayesian regularization prior to increase smoothness of the density estimate over the basketball court and avoid singularities and degeneracies in maximization of the likelihood. 
This can be achieved by adopting the approach of \citet{Fraley:Raftery:2007a}, who proposed weekly informative conjugate priors to regularize the estimation process. 
The EM algorithm can still be used for model fitting, but maximum likelihood estimates (MLEs) are replaced by maximum a posteriori (MAP) estimates. 
For details see \citet[Sec. 7.2]{mclust:book:2023}.

A crucial aspect in mixture modeling is the choice of the number of mixture components, $G_k$, for each outcome. Typically, the Bayesian Information Criterion \citep[BIC;][]{Schwarz:1978} is used as model selection criterion in finite mixture models. 
This choice is justified by \citet{Keribin:2000}, who demonstrated that BIC is consistent for choosing the number of components in a mixture model, assuming a bounded likelihood (which is guaranteed by the introduction of the regularized prior mentioned earlier). 
However, when Bayesian regularization is introduced a slightly modified version of BIC should be used for model selection, with the maximized log-likelihood replaced by the log-likelihood evaluated at the MAP.

\section{Applications}
\label{sec:applications}

In this section we analyzed the player-by-player data of some selected NBA players for the 2022-23 NBA regular season. The data are obtained from the \texttt{R} package \code{hoopR} \citep{Rpkg:hoopR}, which provides easy access to data available on ESPN analytics at \url{https://www.espn.com/nba/}.

\subsection{Stephen Curry}

\begin{figure}[htb]
\centering
\includegraphics[width=\textwidth]{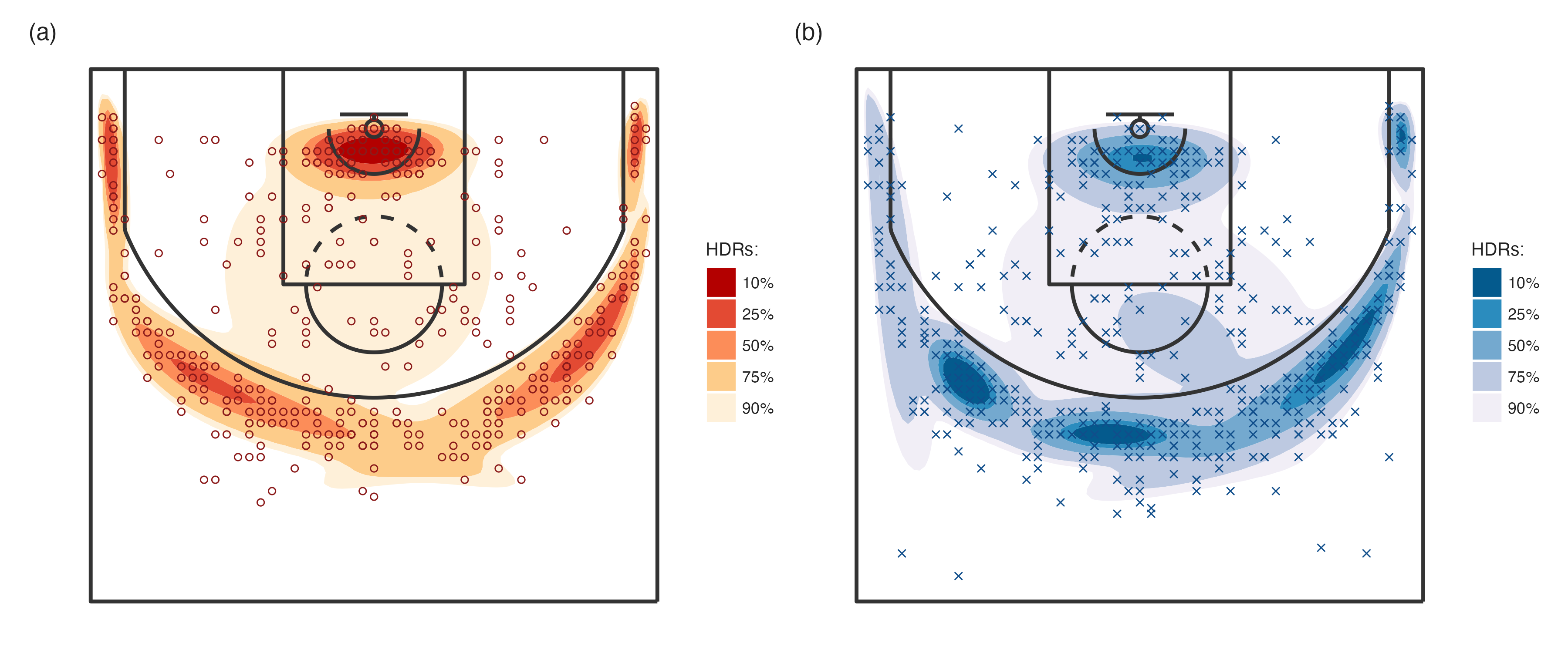}
\caption{Highest density regions (HDRs) from mixture-based estimated densities for made (a) and missed (b) shots for Stephen Curry during the 2022-23 NBA regular season.}
\label{fig:curry-made_miss}
\end{figure}

Figure~\ref{fig:curry-made_miss} shows the estimated densities for made (a) and missed (b) shots, respectively $f(x,y|C=1)$ and $f(x,y|C=0)$ from \eqref{eq:xydens}.
Regions are highlighted by highest density regions (HDRs) corresponding to specific percentages of the data. Note, however, that these cannot be directly compared, but they can be used for computing shot probabilities using \eqref{eq:xycondprob}. 
Required prior probabilities are estimated using proportions of made and missed shots during the regular season, giving $\hat{\tau}_1 = 0.4724$ and  $\hat{\tau}_0 = 0.5276$.

\begin{figure}[htb]
\centering
\includegraphics[width=\textwidth]{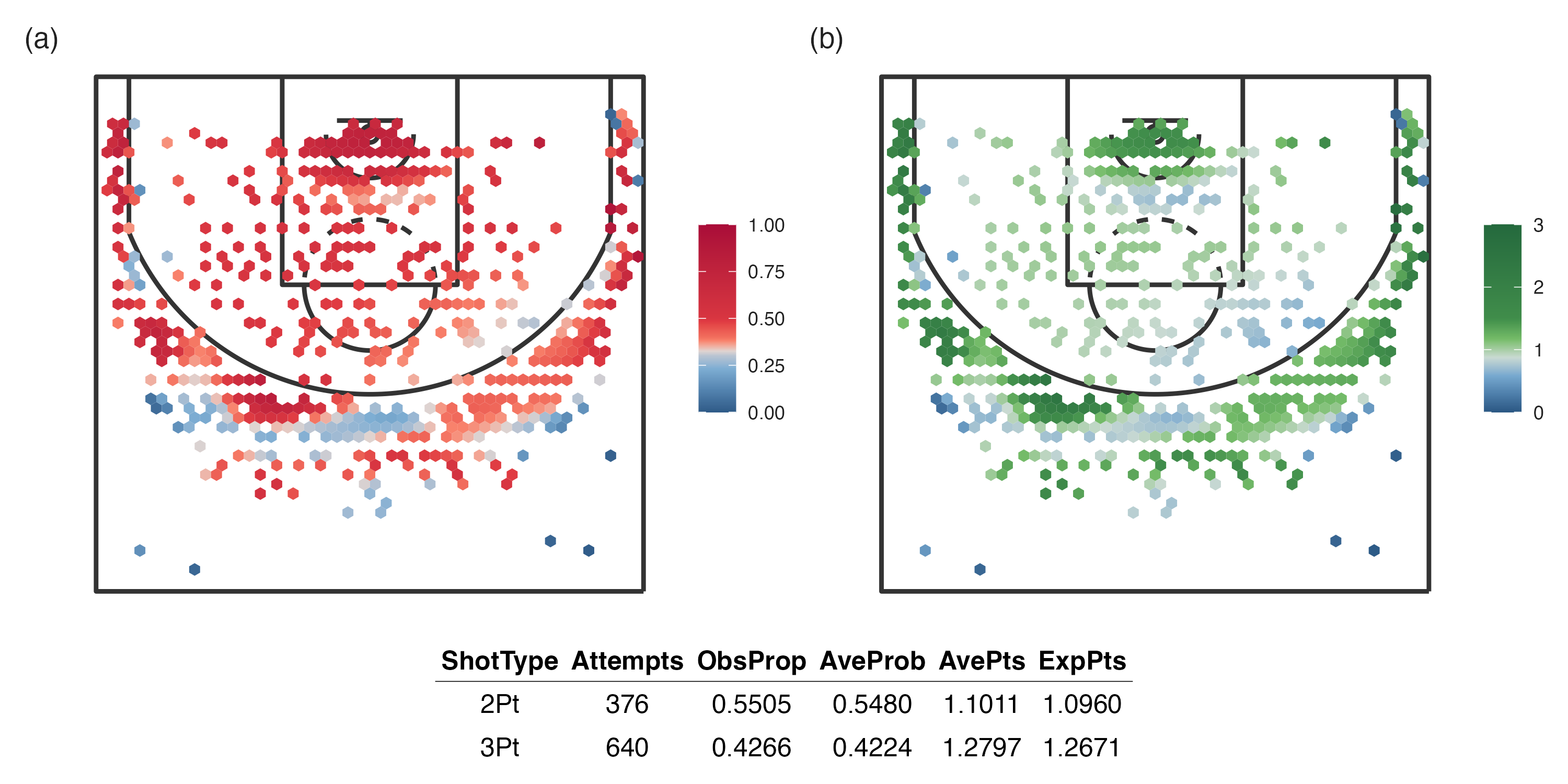}
\caption{Shot charts depicting (a) estimated probabilities and (b) expected points scored per attempt for Stephen Curry during the 2022-23 NBA regular season. The table below the charts reports a summary of empirical and estimated key statistics for both two-point and three-point shots.}
\label{fig:curry-shotcharts}
\end{figure}

Figure~\ref{fig:curry-shotcharts}a presents the estimated shot chart highlighting regions of high and low probability for made shots by Stephen Curry. The chart reveals a remarkable consistency in Curry's shooting ability across various regions, with particularly high probabilities close to the basket and extending well beyond the three-point line. Notably, two key exceptions emerge: very far locations and positions approximately 2-3 feet from the three-point line at the top of the key. Additionally, a closer look suggests a reduced probability in the right mid-range area.

Building upon the estimated shot chart discussed above, Figure~\ref{fig:curry-shotcharts}b presents the corresponding graph of expected points scored. This visualization highlights regions of high scoring efficiency, primarily concentrated around close-range shots and extending to all areas beyond the three-point arc, with a notable preference for the left side. Interestingly, these high-efficiency regions align with areas of higher shot probability observed in Figure~\ref{fig:curry-shotcharts}a, while regions with lower expected points coincide with areas of lower shot probability.

Lastly, the table below Figure~\ref{fig:curry-shotcharts} summarizes key statistics for both two-point and three-point attempts: number of attempts, observed made shot proportions, estimated average probabilities, observed average points per attempt, and estimated expected score. Notably, the empirical and estimated values exhibit close agreement, highlighting the accuracy of the model. These data showcase Stephen Curry's remarkable offensive efficiency beyond the three-point arc, reflected in an estimated expected score of 1.27 points per attempt compared to 1.10 points for closer shots.

\subsection{Joel Embiid}

As a second player we analyze Joel Embiid of the Philadelphia 76ers. Compared to Stephen Curry's role as shooting guard, Embiid plays as a center, is much taller and stronger physically, but at the same time has an excellent aptitude for shooting from mid-range and beyond the arc. During the 2022-23 regular season Embiid had the highest average points per game (30.6) and won the MVP award.

\begin{figure}[htb]
\centering
\includegraphics[width=\textwidth]{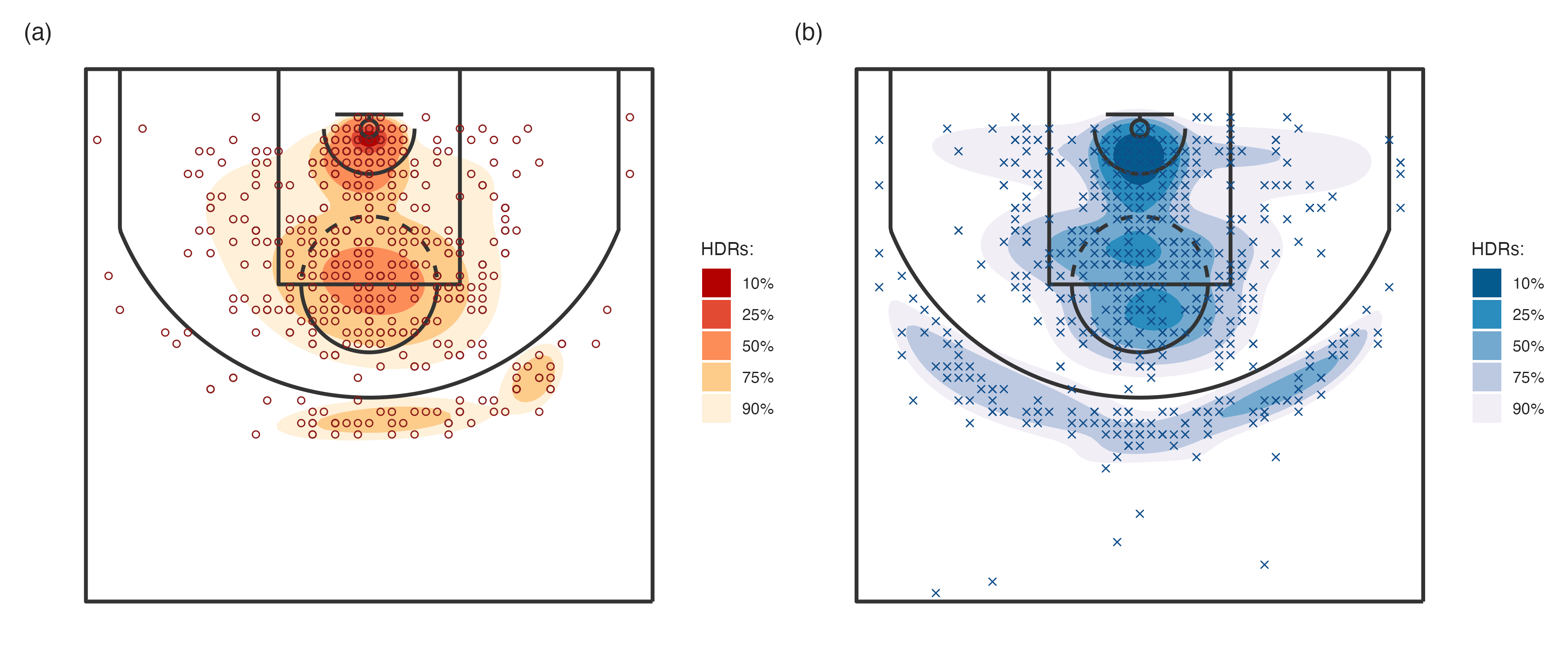}
\caption{Highest density regions (HDRs) from mixture-based estimated densities for made (a) and missed (b) shots for Joel Embiid during the 2022-23 NBA regular season.}
\label{fig:embiid-made_miss}
\end{figure}

Charts in Figure~\ref{fig:embiid-made_miss} show the highest density regions (HDRs) obtained from mixture-based estimated densities for made (a) and missed (b) shots. 
The majority of shots are concentrated in the paint and near the free-throw line, while beyond the three-point arc Embiid's favorite position appears to be the central one.

\begin{figure}[htb]
\centering
\includegraphics[width=\textwidth]{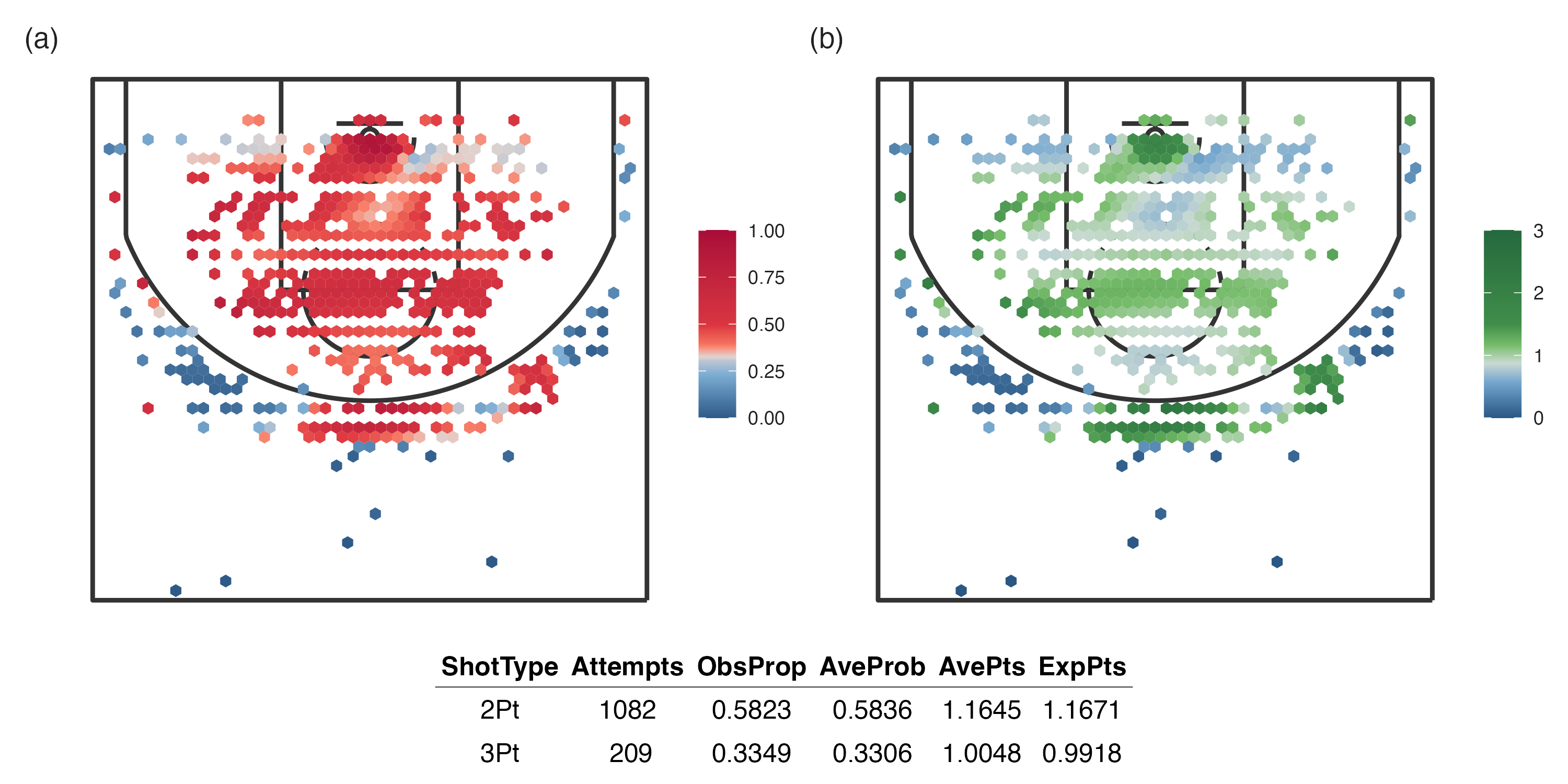}
\caption{Shot charts depicting (a) estimated probabilities and (b) expected points scored per attempt for Joel Embiid during the 2022-23 NBA regular season. The table below the charts reports a summary of empirical and estimated key statistics for both two-point and three-point shots.}
\label{fig:embiid-shotcharts}
\end{figure}

Embiid's shooting efficiency is very high, as can be seen from the chart in Figure~\ref{fig:embiid-shotcharts}a, with estimated success probabilities well above 50\% in almost all mid-range and close-to-basket positions. For three-point shots, two preferred positions with very high success rates emerge: in front of the basket and slightly to the right. In other positions beyond the arc, the estimated probabilities appear significantly lower.

In terms of expected points scored from different positions, the most profitable ones are near the basket, thanks to the high shooting percentage, and those with the highest efficiency beyond the arc, due to the fact that more points are obtained for each basket made.

\medskip

Finally, it is interesting to compare the different shooting choices of Stephen Curry and Joel Embiid, and their relative effectiveness and efficiency (see tables at the bottom of Figures \ref{fig:embiid-shotcharts} and \ref{fig:curry-shotcharts}).
Curry favors long-distance shots, attempting approximately 70\% more three-point shots (640 attempts compared to 376), while Embiid notably focuses on within the arc shots (1082 attempts compared to 209). Curry exhibits high estimated probabilities of scoring for both 2-point (54\%) and, especially, 3-point (42\%) shots, whereas Embiid demonstrates a higher percentage in the mid- and close-range shots (58\%), but only a moderate 3-point percentage (33\%), which is nonetheless excellent for his role. These translate into excellent expected points for both 2-point and 3-point attempts, with Curry astonishingly averaging about 1.27 points per 3-point attempt.


\section{Conclusions}

The availability of good quality spatial data in sports has increased a lot their usage, including spatial visualizations. For example, we are all familiar with heatmaps that represent the location density of players as an attempt to describe their playing behavior but also to identify tactics.
Shot charts are pivotal tools in basketball analytics, offering valuable insights into players' shooting tendencies and efficiencies across different areas of the court. 
Existing shot chart representations often fall short in accurately capturing  shooting spatial distribution, primarily due to their inability to account for the bounded nature of the basketball court.
In the present paper we proposed a new approach that employs Gaussian mixtures to estimate the density distribution using a transformation-based approach that takes into account the physical boundaries of the court.
We demonstrate the effectiveness of our methodology through case studies involving real-world data from the 2022-23 NBA regular season.
 
The easiness of applying and fitting Gaussian mixtures to estimate the spatial distribution creates additional opportunities. 
An explicit extension of the proposed work relates to all other sports where spatial location data are used. Recall also that this  may extend to other non-sport related applications where boundaries need to be taken into account. 
As a proposal for further investigation, we also mention the use of mixture models as the basis for \emph{conditional heatmaps}. 
So far, most of the sports visualization based on tracking data is based on the position of a player in the court. Sometimes it is interesting to visualize the conditional heatmap, i.e. the position of a player conditional on the position of some other player. For example, in basketball (but also in football and other team sports) this can reveal important tactical aspects and space creation strategies for the teams, which is an important ingredient of the game. 
Gaussian mixtures allow easily to work on that since one can easily obtain/estimate the joint distribution of the location of two players as the joint distribution in 4 dimensions, allowing also for dependence. From the joint distribution one can estimate the conditional density in a straightforward manner and thus produce a conditional heatmap.

\clearpage
\bibliography{paper_arxiv}

\end{document}